# On The Feasibility of a Pulsed 14 TeV c.m.e. Muon Collider in the LHC Tunnel

D. Neuffer,[*] V. Shiltsev

  *Fermilab, P. O. Box 500, Batavia IL 60532, USA*
  *E-mail*: neuffer@fnal.gov

ABSTRACT: We discuss the technical feasibility, key machine parameters and major challenges of a 14 TeV c.m.e. muon-muon collider in the LHC tunnel. The luminosity of the collider is evaluated for three alternative muon sources – the PS synchrotron, one of a type developed by the US Muon Accelerator Program (MAP) and a low-emittance option based on resonant muon pair production. Project affordability is also discussed.

---

[*] Corresponding author.

# Contents

## 1. Introduction

A next generation energy-frontier particle physics facility must provide an energy reach beyond that of the LHC, with the potential for the discovery of new physics, and still be affordable within future available budgets [1,2,3]. The proposed pulsed 14 TeV c.m.e. muon-muon collider in the CERN's 27 km tunnel [4] – see Fig.1 - will have a significant (factor of 6-10) advantage in energy reach compared with the existing proton-proton LHC and, therefore, outstanding discovery potential, despite somewhat lower luminosity [5]. The 0.146s lifetime of a 7 TeV muon enables storage and collisions for thousands of turns; that is a great advantage over the single turn of useful collisions possible in a light lepton ($e^+$-$e^-$) collider [6]. The collider cost is expected to be feasible because of the re-use of existing tunnels and the CERN injection complex, as well as the use of cost-efficient magnets and a very limited use of expensive SRF acceleration [7, 8]. The other expected advantages of this collider are a narrow c.m. energy spread in collisions and an outstanding energy efficiency (luminosity per MW of wall-plug electric power) [9, 10, 11].

## 2. Acceleration

Due to their limited lifetime ($\tau = 2.2\gamma$ μs), the acceleration of muons must be fast, so that the number of surviving muons $N_f = N_0(\gamma_0/\gamma_f)^k$, where $k=(0.105\ \text{GeV}/\Delta E)(C/660\ \text{m})$, will be acceptable. That requires a high energy gain $\Delta E$ per revolution of circumference $C$ and a correspondingly fast change in the average magnetic field $<B> = 2\pi E/(0.3C)$. We assume that the LHC tunnel contains a rapid cycling synchrotron (RCS) ring which is filled with a combination of short and very high field SC magnets with DC fields of $B_{SC}$ and longer and weaker pulsed magnets that change their fields from $-B_{pls}$ to $+B_{pls}$ (see Fig. 2). For the ratio of the fields $f = B_{SC}/B_{pls}$ and the required range of acceleration $R= E_{max}/E_{inj}$, the ratio of the lengths of these magnets is $L_{pls}/L_{SC}=f(R-1)/(R+1)$ and the maximum attainable beam energy is equal to $E_{max} = B_{SC}\times(\Pi C/\pi)\times 0.3/(1+f+(1-f)/R)$, where $\Pi < 1$ is the magnet packing factor (ratio of the total magnet length to the ring circumference).

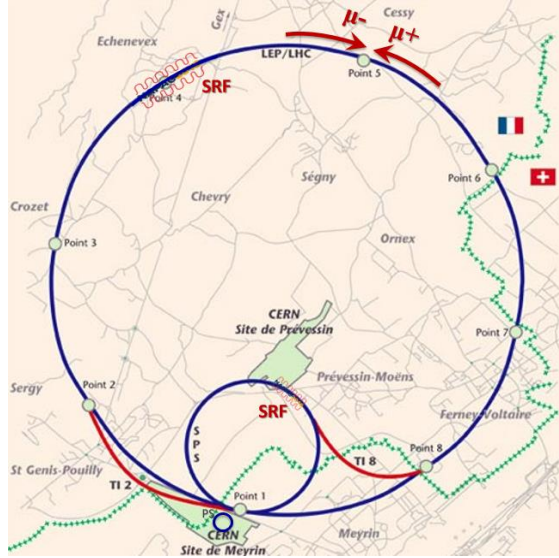

Figure 1: Schematic layout of a pulsed 14 TeV c.m.e. muon collider in the LHC tunnel (7 TeV beam energy).

The optimum choice of the accelerator magnet parameters depends on the technology limits for the SC and pulsed magnets. Table 1 presents key parameters for the accelerators under the assumptions of $\Pi=0.85$, a 50% muon survival per stage ($N_f/N_0=0.5$) and the availability of 16 T $Nb_3Sn$ SC magnets in the LHC tunnel and 8 T NbTi SC magnets in the SPS tunnels.

Beam acceleration from 0.45 TeV to 7 TeV can be done either in a single stage using 3.8T pulsed magnets, or – if the maximum pulsed field is limited to 2 T – in two stages (see the options "LHC-S" and "LHC-D" in Table 1). We use 16 T SC magnets, which are actively and successfully being developed for the Future Circular Collider (FCC) project [12]. The required pulsed magnets could either be superconducting or normal-conducting - up to 5T peak fields have been demonstrated in ~2ms pulsed prototypes [13-16]. The former are more economical. In spite of a number of specific issues, such as AC loss, cooling, quench detection and protection, field quality and material fatigue [17], SC ramping rates of ~1000 T/s are believed to be achievable in HTS-conductor based super ferric magnets [18-20]. Table 1 also includes acceleration parameters of the 30—450 GeV accelerator option "SPS" located in the SPS tunnel that accelerates the muons to the injection energy of the LHC size ring.

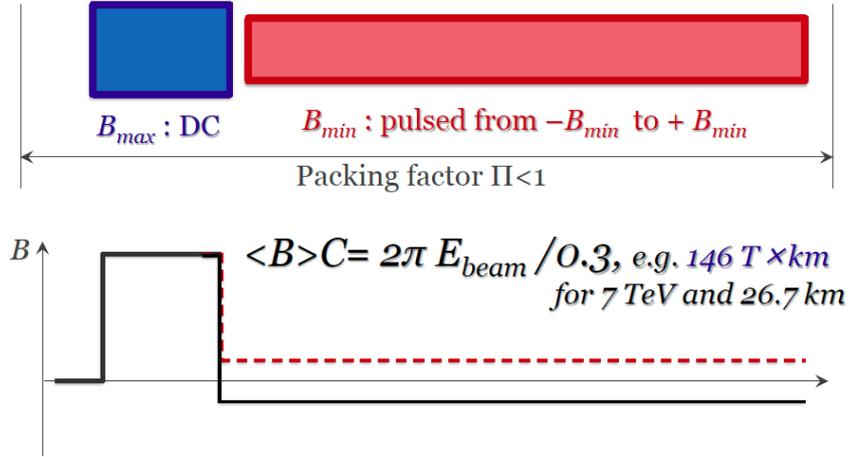

Figure 2: Arrangement (top) and field strengths (bottom) in the SC and pulsed magnets of the pulsed muon rapid cycling synchrotron (RCS).

An alternative acceleration scheme is a recirculating linac (RLA), in which the arcs could be composed of multi-pass (5-20 turns) non-scaling FFAG lines [21], similar to those proposed for eRHIC in [22] and being constructed for the CBETA test accelerator [23]. These would be non-ramping fixed-field magnets, e.g. up to 8T NbTi SC magnets. With fixed-frequency rf, a continuous train of bunches could be acceleraterated to full energy for accumulation in the collider ring. Non-scaling or recently proposed scaling FFAG arcs [24] could also be considered in multi-pass synchrotron scenarios (100-1000 turns) which would use ramped frequency RF acceleration systems.

**Table 1: Muon RCS Accelerator Parameters**

| Scenario | "LHC-S" | "LHC-D" | "SPS" |
|---|---|---|---|
| $C$, km | 26.7 | 26.7 | 26.7 | 6.9 |
| $E_{max}$, TeV | 7 | 7 | 4 | 0.45 |
| $E_{inj}$, TeV | 0.45 | 4 | 0.45 | 0.03 |
| $f_{rep}$, Hz | 5 | 4 | 4 | 20 |
| $\Delta E$/turn, GeV | 14.0 | 3.5 | 9.2 | 3.7 |
| $B_{SC}$, T | 16 | 16 | 16 | 8 |
| $L_{SC}$, km | 4.8 | 7.1 | 2.9 | 0.63 |
| $B_{pls}$, T | 3.8 | 2.0 | 1.9 | 0.8 |
| $\tau_{ramp}$, ms | 42 | 76 | 34 | 2.6 |
| $dB_{pls}/dt$, T/s | 180 | 52 | 112 | 615 |

## 3. Cooling and Luminosity

The collider performance is determined by the intensity and brightness of the muon source. Table 2 summarizes key parameters of three options for a 14 TeV muon collider in the LHC tunnel (i.e., circumference of 26.7 km, beam energy of 7 TeV and beam lifetime of 0.146 s): the first one adapts the existing 24 GeV CERN PS source, the second requires a new 8 GeV linac plus storage ring while the third is based on a threshold $\mu^+$-$\mu^-$ low-emittance muon collider (LEMC) source [25, 26, 27].

### 3.1 PS and MAP options

The first two scenarios are extensions of the US Muon Accelerator Program (MAP) work, which explored the feasibility of a muon collider, and developed detailed scenarios for 1.5, 3.0 and 6.0 TeV colliders [28, 11].

The key components of a MAP collider system are displayed in block diagram form in Figure 3(a): a high-intensity proton source, a multi-MW target and transport system for $\pi$ capture, a front-end system for bunching, energy compression and initial cooling of $\mu$'s from $\pi$ decay, muon cooling systems to obtain intense $\mu^+$ and $\mu^-$ bunches, acceleration up to multi TeV energies, and a collider ring with detectors for high luminosity collisions. The parameters in Table I are scaled from the 6 TeV collider scenario presented in [11]. In the PS case, the ring cycles at 1.2 Hz, accelerating 8 bunches of $6·10^{12}$ protons ($4.8·10^{13}$ total), which is a modest extrapolation of achieved parameters. A fixed energy storage ring could place one compressed bunch at a time (5 Hz) onto a target for muon production and cooling following the baseline MAP scenario. This low-cost scenario would be an order of magnitude lower power than the MAP case, with consequently lower luminosity. It could be a first-stage scenario, to be upgraded later by a new high-power proton source.

The baseline MAP proton source produces pulses of $2·10^{14}$ 8 GeV protons at 15 Hz. This is scaled back to 5 Hz to match the 7 TeV beam lifetime and the above-considered RCS cycle time. The MAP scenario also requires a km-scale cooling system with high-field magnets and RF, and still obtains a relatively large emittance beam for the collider.

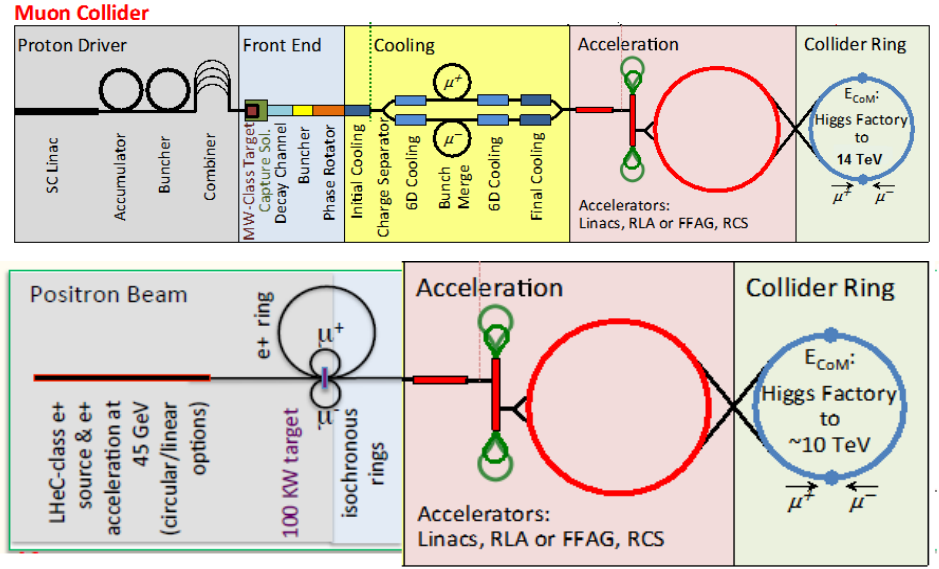

Figure 3: (a) Top- block diagram of a muon collider facility, as studied under MAP [11]; (b) bottom – same for the LEMC option [27].

Table 2: Options for a 14 TeV $\mu^+$-$\mu^-$ Collider

| Parameter | "PS" | "MAP" | "LEMC" |
|---|---|---|---|
| Avg. luminosity | $1.2 \cdot 10^{33}$ | $3.3 \cdot 10^{35}$ | $2.4 \cdot 10^{32}$ |
| Beam $\delta E/E$ | 0.1% | 0.1% | 0.2% |
| Rep rate, Hz | 5 | 5 | 2200 |
| $N_\mu$/bunch | $1.2 \cdot 10^{11}$ | $2 \cdot 10^{12}$ | $4.5 \times 10^7$ |
| $n_b$ | 1 | 1 | 1 |
| $\varepsilon_{t,N}$ mm-mrad | 25 | 25 | 0.04 |
| $\beta^*$, mm | 1 | 1 | 0.2 |
| $\sigma^*$(IR), μm | 0.6 | 0.6 | 0.011 |
| Bunch length, m | 0.001 | 0.001 | 0.0002 |
| $\mu$ production source | 24 GeV $p$ | 8 GeV $p$ | 45 GeV $e^+$ |
| $p$ or $e$/pulse | $6 \cdot 10^{12}$ | $2 \cdot 10^{14}$ | $3 \cdot 10^{13}$ |
| Driver beam power | 0.17MW | 1.6MW | 40 MW |
| Acceleration, GeV | 1-3.5, 3.5-7 RCS | 1-3.5, 3.5-7 RCS | 40 GV, RLA 20 turn |
| $v$ radiation, mSv/yr | 0.05 | 0.8 | 0.008 |

### 3.2 Low Emittance Muon Collider (LEMC) option

Recently, a scenario that uses resonant production of $\mu^\pm\mu^-$ pairs at threshold from $e^+e^-$ collisions has been developed [25, 26, 27]. This could have the advantage of producing $\mu^\pm\mu^-$ with very small emittances; it has the disadvantage of low production rate. Fig. 3(b) shows an overview of the scenario. The primary engine for this source is a ~45 GeV positron storage ring. Bunches of positrons collide with an electron target (within a material slab) producing $\mu^+$ and $\mu^-$ at threshold, each with ~22 GeV/c momentum. The small transverse-momentum at threshold production and small spot size with small positron beam emittance creates μ beams with small emittance. With a slab target of 0.3mm Be, ~$10^{-7}$ muon pairs per $e^+$ bunch pass are obtained. A 6.3 km circumference ring with 100 bunches of $3\times10^{11}$ $e^+$ feeds 63 m circumference 22 GeV $\mu$ rings, which accumulate muons for ~ 2500 turns, obtaining bunches of ~$4.5\times10^7$ muons at ~2200Hz ($10^{11}$ $\mu$/s) [27]. The $e^+$ storage ring in this scenario is quite challenging. At $3\cdot10^{13}$ $e^+$ stored, it produces ~140MW of synchrotron radiation. An $e^+$ lifetime of ~250 turns implies a beam source of 40 MW of 45 GeV $e^+$ is required and ~$5\cdot10^{15}$ $e^+$/s, much larger than that readily available from modern day positron sources. Beam dynamics simulation [28] showed lifetimes of only ~40 turns in a model lattice; this would then require ~$3\cdot10^{16}$ $e^+$/s (250 MW). The low emittances shown in Table 1 only includes the effect of the muon production divergence and scattering in a single pass through the target; a future evaluation in a complete scenario is needed, which would include positron beam and multipass effects.

The scenario accumulates muons at 22 GeV, and therefore has a natural cycle time of ~0.45 ms (~2.2 kHz). A special type of fast RLA accelerator will be required, e.g. 40 GeV/turn in the LHC tunnel providing 7 TeV of energy gain in 170 turns (15 ms). A CW SRF RLA system should be capable of accelerating simultaneously 15 ms/ 0.45 ms=34 bunches. Then the bunches are delivered to a 26.7 km 7 TeV DC collider ring at 2200 Hz, where each bunch collides at the detector for the muon lifetime of 1600 turns. At any given time ~200 bunches would be colliding, though with relatively low average luminosity as indicated in Table 2. The luminosity could be increased if the bunches could be superimposed, however, phase space conservation implies that bunch combination will increase emittance, reducing luminosity. A scenario that increases net luminosity must be identified.

### 3.3 Neutrino radiation considerations

A potential limitation in high energy muon colliders is the long-range radiation from the neutrinos produced from muons that decay in the Collider ring. Because of the high energy of the muons, the resulting neutrinos are emitted in a narrow plane along the ring orientation ($\theta \cong m_\mu/E_\mu$) and interact at a rate proportional to $E_\mu^2$. While the interaction rate is low, it may accumulate into a radiologically significant dose where the beam reaches the earth's surface. Approximate formulae for that radiation dose in an idealized ring can be obtained from Refs. [29, 30, 31, 32]. We use the methods presented in references [30, 31] to obtain :

$$Dose \cong 1.5 \frac{N_\mu' E_\mu^3}{R_x^2} mSv/year$$

where $N_\mu'$ is the number of muons/second (in $10^{13}$/s), $E_\mu$ is in TeV and $R_x$ is the distance from the ring to surface exit in km. $R_x$ is 36 km for a 100m deep ring in an idealized geometry, and an operational year of $10^7$ s is used. As discussed in ref. 31, the formula should be a conservative overestimate, since it assumes that all muons decay in the ring, the width of the plane of emission is not increased by the finite sizes of the beam emittance and of the radiation showers, and the cross sections increase proportional to $E_\nu$. (Above ~1TeV, the neutrino cross section increase is reduced toward ~ $E_\nu^{1/2}$.) It also uses an immersion model which enhances exposure over that in realistic occupancy models.

This obtains the estimates shown in Table 2, where we have added contributions from both muon beams, conservatively assuming their planes of emission precisely overlap. The CERN external control limit is ~1 mSv/yr [32] and the MAP case is near that level. Some mitigation may be required. A more complete evaluation is needed. The radiation density can be reduced by about an order of magnitude by adding a vertical orbit variation of a few mm.

The lower luminosity PS and LEMC examples are relatively safe. The factor of 100 fewer muons used in the LEMC design provides a large margin of safety and is an important advantage of that scenario.

## 4. Discussion of Financial Feasibility

It is foreseeable that a 14 TeV lepton collider could have a physics reach greater than a 100 TeV hadron collider. A critical difficulty toward the feasibility of performance is the muon beam cooling scenario. The MAP collaboration developed cooling scenarios that cool muons taken from a production target from $\varepsilon_t$ ~0.02 m-rad to ~25 μm as well as adequate designs of very small β* beam optics – see the series of articles in a JINST Special Issue [33]. The first successful muon ionization cooling results are being reported by the MICE collaboration [34]. The "no-cooling" (positron ring based) design needs much more optimization to ease the facility power requirements, and will require further design to establish low emittances in the multi-turn muon accumulator ring.

Acceleration based on pulsed and CW SRF should generally be considered feasible for gradients of about 30 MV/m (pulsed) and 20MV/m (CW). The required pulsed magnets exist only in prototypes and significant technology development is very much needed to prove technical feasibility.

The main attraction of the 14 TeV $\mu^+$-$\mu^-$ collider discussed above is its cost feasibility – see Fig.4. The total project cost (TPC, often cited as the "US accounting" and usually a factor of 2-2.4 greater than the "European accounting") of a future high-energy collider can be roughly estimated according to Eq. (2.1) in Ref. [7] which accounts for civil construction, technical components like normal-conducting and superconducting magnets and superconducting RF, respectively, and for the total required site "wall-plug" power.

That equation is:

$$TPC \cong \alpha \left(\frac{L}{10km}\right)^{1/2} + \beta_M \left(\frac{E}{1TeV}\right)^{1/2} + \beta_{RF} \left(\frac{E_{RF}}{1TeV}\right)^{1/2} + \gamma \left(\frac{P}{100MW}\right)^{1/2}$$

where L is the tunnel length, E is cm energy, $E_{RF}$ is total RF voltage, and P is total site power. α and γ are ~2 B$, and β is ~ 1 B$ for normal conducting magnets, 2 B$ for SC magnets, 8 B$ for normal conducting RF and 10 B$ for superconducting RF.

In our proposal the civil construction costs can be reduced by reusing the existing 27 km LHC tunnel, the 7 km SPS tunnel and the accompanying CERN infrastructure. The incremental cost to build the proposed collider in its least expensive proton source configuration "PS" and combined system of SC and pulsed magnets to get to 7+7=14 TeV using up to 20 GeV of the SRF acceleration would be about 2B$×sqrt(14TeV) + 10B$× sqrt(0.02TeV) = 8.9±3 B$.

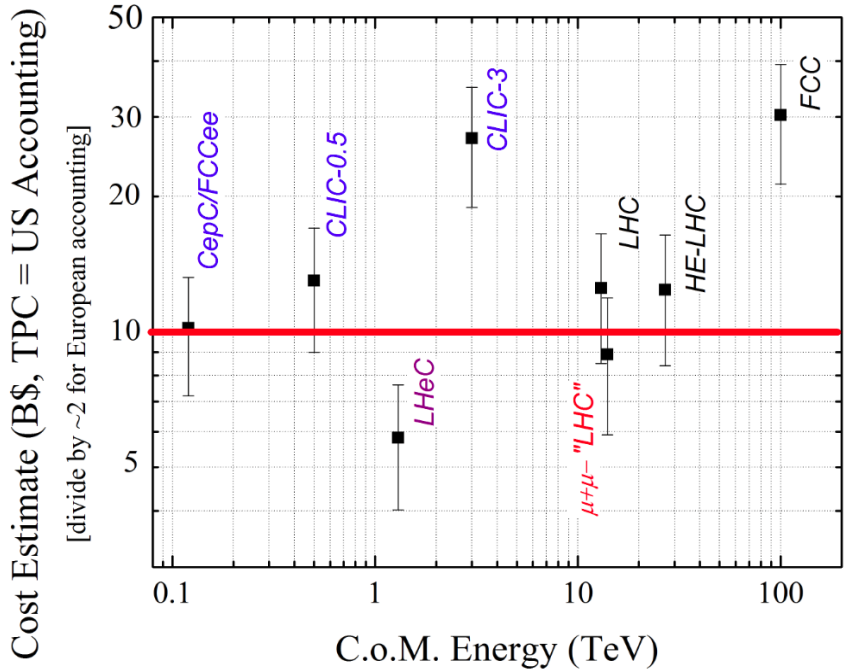

Figure 4: Cost estimates of various future colliders, including a cost estimate for a baseline muon collider in the LHC tunnel (μ+μ-"LHC").

A high power proton driver (2 MW 8 GeV beam, about 20 MW of site power) is needed in the high-luminosity "MAP" configuration - see Table 2 - and will cost an extra ~10B$×sqrt(0.008)+ 2B$×sqrt(0.2)=1.8±0.6 B$.

The most expensive option is the "LEMC" – even if the SPS tunnel is reused, a new 45 GeV positron ring with 120 MW of SR power requiring 1 GV of SRF (some 250MW of total wall plug power) will cost an additional ~1B$×sqrt(0.045)+ 10B$×sqrt(0.001)+2B$×sqrt(2.5) = 3.6 ± 1.2 B$.

Figure 4 shows the least expensive option for a 14 TeV LHC-based muon collider with its error bars (~9±3 B$). The other cases would add only ~1.8—3.6 B$ and not change the scale of the comparisons. Establishment of feasibility along with more detailed cost estimates will be needed to determine the best of these approaches to a future collider.

## Acknowledgments

We thank D. Summers, M. Palmer, F. Zimmermann for helpful discussions and suggestions. This research is supported by Fermi Research Alliance, LLC under Contract No. DE-AC02-07CH11359 with the U.S. Department of Energy, Office of Science, Office of High Energy Physics.